\documentclass[fontsize=11pt]{arxiv-article}
\pdfoutput=1
\bibliography{not-arxiv,References}

\usepackage{Definitions}

\usepackage{hyperref}

\allowdisplaybreaks

\newcommand{\OfficialTitle}{
  2D CFTs -- Large Charge is not enough
}
\title{\setstretch{1.4}
  {\color{Thoughtless}\OfficialTitle}
}

\hypersetup{pdfauthor={Thiago Araujo, Omar Celikbas, Domenico Orlando, Susanne Reffert},pdftitle={\OfficialTitle},%
  colorlinks=true,linkcolor=ThoughtYouWere,citecolor=ThoughtYouWere,urlcolor=ThoughtYouWere,ocgcolorlinks}

\author{%
  \begin{minipage}{.94\textwidth}
    \vspace{1cm}
    \begin{center} \dosserif%
      {\small
      	\textbf{Thiago Araujo}\textsuperscript{\ding{73}}, 
      	\textbf{Omar Celikbas}\textsuperscript{\ding{73}},
        \textbf{Domenico Orlando}\textsuperscript{\ding{72}\ding{73}}\\ and
        \textbf{Susanne Reffert}\textsuperscript{\ding{73}} 
         }
    \end{center}
    \authorBlock{\ding{73}}{\dosserif{}%
      Albert Einstein Center for Fundamental Physics\\
      Institute for Theoretical Physics, University of Bern,\\
      Sidlerstrasse 5, CH-3012 Bern, Switzerland}
    \authorBlock{\ding{72}}{\dosserif{}%
      INFN sezione di Torino.\\
      via Pietro Giuria 1, 10125 Torino, Italy}
  \end{minipage}
}

\date{}

\begin{document}

\numberwithin{equation}{section}

\begin{titlepage}

  \maketitle

  \thispagestyle{empty}

  \vfill\dosserif{}

  \abstract{\normalfont{}\noindent{}%
    In this note we study two-dimensional \acsp{cft} at large global charge. Since the large-charge sector decouples from the dynamics, it does not control the dynamics and an \acs{eft} construction that works in higher-dimensional theories fails.
    It is however possible to use large charge in a double-scaling limit when another controlling parameter is present.
    We find some general features of the spectrum of models that admit an \acs{nlsm} description in a \acs{wkb} approximation and use the large-charge sector of the solvable $SU(2)_k$ \acs{wzw} model to argue the regimes of applicability of both the large-\(Q\) expansion and the double-scaling limit.
}

\end{titlepage}

\setstretch{1.1}
\tableofcontents

\section{Introduction}%
\label{sec:Introduction}

In more than two dimensions, studying \acp{cft} in sectors of large global charge leads to important simplifications and allows the semiclassical computation of the conformal data as an expansion in inverse powers of the large charge.
The large-charge expansion has been put to work in various dimensions~\cite{Hellerman:2015nra,Gaume:2020bmp,Alvarez-Gaume:2016vff,Monin:2016jmo,Orlando:2019hte,Loukas:2017lof,Arias-Tamargo:2020fow,Cuomo:2020rgt,Moser:2021bes}, but the case of two-dimensional \acp{cft} has received much less attention and has been studied only in~\cite{Komargodski:2021zzy}.
There it was shown that the approach of writing an \ac{eft} -- so successful in higher dimensions -- fails in $D=2$.
If we impose unitarity and discreteness of the spectrum, the \(U(1)\) sector corresponding to the fixed charge can describe only a free boson with central charge $c=1$ and completely decouples from the rest of the theory as a consequence of Sugawara's construction~\cite{Sugawara:1967rw}.
The large charge does \emph{not} control the full low-energy dynamics, unlike in the higher-dimensional case, where the dynamics is controlled by the scale introduced by the chemical potential.
This means that in two dimensions the scaling dimension of the lowest operator of fixed charge $Q$ is given by $Q^2$ (from the $c=1$ free boson) plus contributions not controlled by the charge.

While using the charge as a controlling parameter in an \ac{eft} for an otherwise strongly coupled model fails, we can still use it in regimes where the theory has a controlling parameter of its own.
Despite the failure of the \ac{eft} construction, studying sectors of large charge still allows us to extract general insights about the spectrum. 
If the \ac{cft} can be described by a \ac{nlsm} in a particular limit, working at large charge simplifies the analysis in analogy to the double-scaling limits considered in higher-dimensional theories~\cite{Alvarez-Gaume:2019biu,Orlando:2019hte,Giombi:2020enj,Dondi:2021buw,Grassi:2019txd,Watanabe:2019pdh,Badel:2019oxl,Badel:2019khk,Arias-Tamargo:2019xld,Antipin:2020abu,Antipin:2020rdw,Antipin:2021akb,Jack:2021lja,Jack:2021ypd,Jack:2021aui}.
We find that, in such a regime, generically the conformal dimension of the lowest operator of charge \(Q\) is written as an expansion in $1/Q$ starting at order $\order{Q^2}$.

\bigskip

\noindent{}In this note, we first compute the spectrum of a system with an \ac{nlsm} description using a geometrical approach in which the large charge is the controlling parameter in a \ac{wkb} approximation.
Then, we exploit the fact that some two-dimensional \acp{cft} are exactly solvable to compare our large-charge results with the exact partition function specialized to a sector of fixed charge.
This allows us to verify our general results and spell out the precise regimes of validity of the large-charge expansion and of the double-scaling limit.

We make in particular use of the fact that \ac{wzw} models at level $k$ admit a geometrical interpretation in the limit $k \to \infty$.
For the $SU(2)$ \ac{wzw} model, this limit corresponds to an \ac{nlsm} with target space $S^3$.
This model has a \(SU(2) \times SU(2)\) global symmetry that we can use to fix two independent charges \(Q\) and \(\bar Q\).
In the limit $k \gg Q, \bar Q \gg 1$, we can use the \ac{wkb} approximation and  find the scaling dimension 
\begin{equation}
	\Delta = \frac{(Q+\bar Q)(Q+\bar Q+2)}{2k},
\end{equation}
which is matched by the exact result from the partition function.
Also the marginal $J \bar J$ deformation, where the symmetry is reduced to \(U(1) \times U(1)\), can be treated in the same way.
In this limit, the charges are not the dominating controlling parameter, but still serve to simplify the computation.
Our treatment of the $SU(2)_k$ \ac{wzw} model gives a proof of concept for the usefulness of working at large charge in a double scaling limit together with the controlling parameter of the theory.
This approach will be valuable in the study of more general models for which an exact solution is not known.

\bigskip

\noindent{}The plan of this note is as follows. In Section~\refstring{sec:nlsm} we study \acp{cft} which by assumption have an \ac{nlsm} description. The most general such action is the one of the string worldsheet.
In Section~\ref{sec:cylH}, we make use of some classical string theory results that allow us to identify the operator appearing in the one-loop tachyon beta-function equation with the cylinder Hamiltonian, which in geometrical terms is interpreted as a generalized Laplacian.
In Section~\ref{sec:wkb}, we observe that in the limit of large charge, the eigenvalue equation of this Laplacian has the right form to admit a \acs{wkb} approximation.
In Section~\ref{sec:WKB_Examples}, we consider three examples, in which the \acs{wkb} hierarchy can be solved: the case of two-dimensional target space, the case of the three-sphere corresponding to the semi-classical $k \to \infty$ limit of the $SU(2)_k$ \ac{wzw} model, and the marginal deformation of this latter example.

In Section~\refstring{sec:wzw}, we consider the fixed-charge sectors of the completely solvable $SU(2)_k$ \ac{wzw} model and its marginal deformations starting from the exact partition function.
First we briefly introduce the \ac{wzw} model (Section~\ref{sec:wzw-model}) and the parafermion decomposition (Section~\ref{sec:parafermions}), specializing the general results to a sector of fixed charge and finding the lowest-energy state.
Via the state-operator correspondence, this leads us directly to the scaling dimension of the lowest operator of fixed charge.
Two regimes emerge: if $(Q +\bar Q) < k$, the large charge is not the dominating controlling parameter and we match the \ac{wkb} results from Section~\ref{sec:WKB_Examples} in the limit $k\to \infty$.
For $(Q +\bar Q) > k$, the \(U(1)\) sector decouples and only controls a subsector of the full dynamics.

In Section~\refstring{sec:outlook}, we give brief conclusions and an outlook.
In Appendix~\ref{sec:freeBoson}, we discuss the free boson at large charge.
\section{The non-linear sigma model at large charge}%
\label{sec:nlsm}

We start by considering \acp{cft} for which, by assumption, an \ac{nlsm} description exists.
This is for example the case if a Lagrangian description can be realized in a semi-classical approximation thanks to the existence of a small parameter.
In this case, in string-theoretical language, the \ac{cft} is described in terms of background fields (the metric, the \(B\) field, the dilaton, the tachyon) living on a target space.
In particular, the spectrum of the dilatation operator is identified with the spectrum of a differential operator that is constructed using these background fields.

One possible way of constructing this operator was proposed in~\cite{Tseytlin:1993my,Frohlich:1993es}
in order to realize a geometrical description for a given \ac{cft}.
The idea is to study the beta function of the lowest-lying state in the \ac{nlsm}, \emph{i.e.} the tachyon, in order to identify the cylinder Hamiltonian of the theory with the generalized (string-frame) Laplacian on the target space.

In this section we first summarize this construction and then show how, in presence of a \(U(1)\) global symmetry in the \ac{cft} (which translates into a \(U(1)\) isometry for the target space) the spectrum of this Laplacian can be studied in a \ac{wkb} approximation in the limit of large fixed charge \(Q\).

\subsection{The cylinder Hamiltonian as a differential operator}
\label{sec:cylH}

The most general action with up to two derivatives is the \ac{nlsm} of the closed string worldsheet,
\begin{equation}
	S = \frac{1}{4\pi\alpha'} \int \dd{\tau}\dd{\sigma}  \pqty*{ G_{\mu\nu}(X) \del_{\alpha} X^\mu\del^{\alpha} X^\nu + i B_{\mu\nu}(X)\del_\alpha X^\mu \del_\beta X^\nu \epsilon^{\alpha\beta} 
   + \frac{\alpha'}{2} \Phi(X)R^{(2)} + T },
\end{equation}
where \(\mu = 1, \dots, N\), $G$ is the target space metric, $B$ is the antisymmetric Kalb--Ramond field, $\Phi$ is the dilaton, $T$ is the tachyon and $R^{(2)}$ is the Ricci scalar of the worldsheet.
We assume that the system has a global \(U(1)\) (compact) symmetry that is realized non-linearly as the shift of one of the fields $X^N = \varphi$ as $\varphi \to  \varphi+\varepsilon$.
It follows that $\varphi$ can only appear via its derivatives, and the target-space fields cannot depend on $\varphi$.

If we want the \ac{nlsm} to describe a \ac{cft}, we need to study the \(\beta \)-functions for the fields \(G\), \(B\) and \(\Phi\).
At one loop the vanishing of the conformal anomaly takes the form~\cite{Callan:1985ia}:
\begin{align}
  \beta_{\mu\nu}^G &= R_{\mu\nu} - \frac{1}{4} H_{\mu\nu}^2 + 2\nabla_\mu \nabla_\mu \Phi,  \\
  \beta_{\mu\nu}^B &= \frac{1}{2} \nabla^\lambda H_{\lambda\mu\nu} - \nabla^\lambda \Phi H_{\lambda\mu\nu}, \\
  \beta^{\Phi} &= \frac{N}{6}  + \frac{\alpha'}{2} \pqty*{ - R + \frac{H^2}{12} + 4 (\nabla \Phi)^2 - 4 \Laplacian \Phi},
\end{align}
where \(N\) is the number of two-dimensional fields, \(R\) is the curvature of \(G\), \(H = \dd{B}\), \(\nabla\) is the covariant derivative associated to \(G\), \(H^2_{\mu\nu} = H_{\mu\lambda\rho}H\indices{_\nu^\lambda^\rho}\), \(H^2 = H\indices{_\mu^\mu}\).
We require \(\beta^{G} = \beta^B = 0\).
These conditions imply that \(6\beta^\Phi \) is a constant  \(c\) to be identified with the central charge.
As for the tachyon, the one-loop beta-function is a second-order differential equation for \(T\)~\cite{Tseytlin:1993my},
\begin{equation}\label{eq:betaTachyon}
	\beta^T = \pqty*{-\frac{e^{2\Phi}}{2\sqrt{\det G}}\del_\mu\pqty*{e^{-2\Phi}\sqrt{\det G}G^{\mu\nu}\del_\nu } - \frac{c}{12}} T = 0.
\end{equation}
This operator appears in the equations satisfied by all the small perturbations of the target-space fields around the background values at the fixed point \(\beta = 0\).

In order to compute the conformal dimensions in the \ac{cft} we can apply the state-operator correspondence and identify the dilatation operator in the plane with the Hamiltonian on the cylinder, which in a \ac{cft} is generically given by
\begin{equation}
  H_{\text{cyl}} = L_0 + \bar L_0 -\frac{c}{12}.
\end{equation}
We now need to identify the differential operator corresponding to this Hamiltonian in terms of the background fields of the sigma model.
A possible strategy was proposed in~\cite{Tseytlin:1993my,Frohlich:1993es}.
The idea is that the cylinder Hamiltonian appears in string theory in the Virasoro condition that a state has to satisfy in order to be physical:
\begin{equation}
	\pqty*{L_0 + \bar L_0 -\frac{c}{12}}\ket{\text{phys}} =H_{\text{cyl}} \ket{\text{phys}}= 0.
\end{equation}
The tachyon $T$, being the lowest scalar of the theory, must in particular satisfy this condition,
\begin{equation}
	H_{\text{cyl}} T =0.
\end{equation} 
It is thus natural to identify this Virasoro condition with the vanishing of the one-loop beta function for the tachyon that we have written above and the cylinder Hamiltonian with the operator in Eq.~\eqref{eq:betaTachyon}.
It is worth emphasizing that we will not need to impose the Virasoro condition in the following, since we will not be discussing a string theory. We use it here to identify the correct representation of the Hamiltonian.

$H_{\text{cyl}}$ has a geometrical interpretation in terms of the target-space: we can identify it with a generalized (or string-frame) Laplacian. All together, 
\begin{equation}
	H_{\text{cyl}} = -\frac{1}{2}\Laplacian_\Phi - \frac{c}{12}.
\end{equation}
As an aside, for vanishing dilaton $\Phi =0$, this Hamiltonian describes the propagation of a free particle on the space with metric $G_{\mu\nu}$.
This shows that we are actually studying the homogeneous limit of the \ac{nlsm}, or equivalently, the motion of the center of mass of a closed string.

\subsection{WKB approximation at large charge}%
\label{sec:wkb}

We now specialize to a sector of the theory with fixed and large \(U(1)\) charge $Q$. 
We want to compute the spectrum of the generalized Laplacian $\Laplacian_\Phi$ in this sector, \emph{i.e.} the eigenvalues $E(Q)$ of the equation
\begin{equation}
	\frac{1}{2}\Laplacian_\Phi \Psi_Q + E(Q) \Psi_Q = 0.
\end{equation}
The Laplacian is a second-order differential operator, and we can generically expand it in terms of derivatives \emph{w.r.t} the target-space directions
\begin{equation}
  \frac{1}{2}\Laplacian_\Phi{} = A^{mn}(X)\del_m\del_n{} + B^m(X) \del_m \del_\varphi{} + C(X) \del^2_\varphi{} +  D^m(X) \del_m{} + F(X) \del_\varphi{},
\end{equation}
where \(m = 1, \dots, N -1 \), \(\varphi = X^{N}\) and the functions \(A\), \(B\), \(C\), \(D\) and \(F\) are written in terms of the metric and the dilaton:
\begin{equation}
  \begin{aligned}
    A^{mn} &=  \frac{1}{2} G^{m n} \ , & B^m &=  G^{m \varphi} \ , & C &=  \frac{1}{2} G^{\varphi \varphi} \ , \\
    D^m &= \frac{1}{2} \frac{e^{2 \Phi}}{\sqrt{G}} \del_n( e^{-2 \Phi} \sqrt{G} G^{m n}) \ , &  F &= \frac{1}{2} \frac{e^{2 \Phi}}{\sqrt{G}} \del_n( e^{-2 \Phi} \sqrt{G} G^{\varphi n}) \ .
  \end{aligned}
\end{equation}
The global \(U(1)\) symmetry of the \ac{nlsm} is now an isometry of the metric $G$, so in general $A$, $B$, $C$, $D$ and $F$ depend on $X^m$ and not of \(\varphi\), and the eigenfunctions of the Laplacian take the form
\begin{align}
	\Psi_Q(X^m, \varphi) &= \Psi_Q(X^m) e^{iQ\varphi}, & m &= 1, \dots, N -1.
\end{align}
Our problem then reduces to
\begin{equation}\label{eq:sectrumLapExpand}
	\pqty*{A^{mn}(X) \del_m\del_n{} + \pqty*{i Q B^m(X) + D^m(X)} \del_m{} - Q^2 C(X) + i Q F(X) + E(Q)} \Psi_Q(X^m) = 0 \ .
\end{equation}

We are interested in the scaling properties of the spectrum as function of the charge.
For large $Q$, we can use the \ac{wkb} method, which allows to approximate the solution of a differential equation whose highest derivative is multiplied by a small parameter.
If we divide equation~\eqref{eq:sectrumLapExpand} by $Q^2$, we have the exact form suitable for the approximation in the $Q \gg 1$ limit.

We start with the Ansatz
\begin{equation}
	\Psi_Q(X^m) = \exp( Q \sum_{i = 0}^\infty \frac{S_i(X)}{Q^i}) \ .
\end{equation}
The eigenvalue problem can be decomposed into an expansion in powers of \(Q\), starting from \(Q^2\).
It follows that the energy must have the form
\begin{equation}
  E(Q) = E_2 Q^2 + E_1 Q + E_0 + \dots
\end{equation}
and we obtain a hierarchy for the family of functions \(S_i(X)\), starting with the eikonal (leading order) approximation
\begin{equation}
  A^{mn}(X) \del_m S_0 \del_n S_0 + i B^m(X) \del_m S_0 = C(X) - E_2 \ ,
\end{equation}
which, expressed in terms of the metric, has the form 
\begin{equation}
   G^{mn} \del_m S_0 \del_n S_0 + 2 i G^{m \varphi} \del_m S_0 =  G^{\varphi \varphi} - 2 E_2 
\end{equation}
and does not depend on the dilaton.

\subsection{Examples}
\label{sec:WKB_Examples}

The main result of the previous calculation is that if a \ac{cft} is described by an \ac{nlsm} and it has a global \(U(1)\) symmetry, the conformal dimensions of the lowest operators of fixed charge \(Q\) are given by an expansion in \(1/Q\) starting at order \(Q^2\).
In special cases we can solve the \ac{wkb} hierarchy and extract more information about the system.

\paragraph{Two dimensions.}
In the special case of two fields \(X^\mu = (X, \varphi)\), the \ac{wkb} approximation is particularly simple.
In fact, we can always rewrite the spacetime metric as \(G_{\mu \nu} = e^{2 f(X)} \delta_{\mu\nu}\), so that the generalized Laplacian is
\begin{equation}
  \frac{1}{2} \Laplacian_\Phi = \frac{1}{2} e^{-2f(X)} \pqty*{\del^2_X{} -2 \del_X \Phi \del_X{} + \del^2_\varphi {}}.
\end{equation}
Then the \ac{wkb} hierarchy becomes
\begin{equation}
  \begin{aligned}
    S_0'(X)^2 &= 1 - 2 e^{2f(X)} E_2, \\
    S_1'(X) &= \frac{2 S_0'(X) \Phi'(X) - S_0''(X) -2 e^{2f(X)} E_1}{2 S_0'(X)}, \\
    & \vdots
  \end{aligned}
\end{equation}
which can be solved order-by-order
\begin{equation}
  \begin{aligned}
    S_0(X) &= \pm \int^X \dd{\xi} \sqrt{1 - 2 e^{2f(\xi)} E_2},  \\
    S_1(X) &= \Phi(X ) - \frac{f(X)}{2} + \frac{1}{2}  \int^X \dd{\xi}\frac{f'(\xi)}{1 - 2 E_2 e^{f(\xi)} } \pm \frac{E_1 e^{2 f(\xi )}}{\sqrt{1 - 2 E_2 e^{f(\xi)} }}, \\
    & \vdots
  \end{aligned}
\end{equation}
where, at the fixed point, the functions \(f(X)\) and \(\Phi(X)\) are
\begin{align}
  f(X) &= c_1 - \frac{1}{2} \log( 1 - c_2 e^{  c_3 X} ) \ , & \Phi(X) &= c_4 + \frac{c_{3}}{2} X  - \frac{1}{2} \log( 1 - c_2 e^{ c_3 X} ) \ ,
\end{align}
with \(c_i\) being constants.

\paragraph{The three-sphere.}

Another interesting example is the three-sphere \ac{nlsm} that describes the semiclassical \(k \to\infty\) limit of the \(SU(2)_k\) \ac{wzw} model. 
Using the \ac{wkb} approximation we can study the regime \(k \gg Q \gg 1\).
It is convenient to pick a coordinate system in which the two \(U(1)\)s are manifest, for example by embedding the three-sphere in \(\setC^2\) as follows (Hopf coordinates):
\begin{equation}
  \begin{cases}
    z_1 = \rho e^{i \vartheta}, \\
    z_2 = \sqrt{1 - \rho^2} e^{i \varphi }.
  \end{cases}
\end{equation}
The corresponding line element is
\begin{equation}
  \dd{s}^2 = k \bqty*{ \frac{\dd{\rho}^2}{1- \rho^2} + \rho^2 \dd{\vartheta}^2 + \pqty*{1- \rho^2} \dd{\varphi}^2}.
\end{equation}
The Laplacian reads
\begin{equation}\label{eq:Laplacian}
  \Laplacian = \frac{1}{k} \bqty*{\pqty*{1 - \rho^2} \del_\rho^2{} + \frac{1 - 3 \rho^2}{\rho} \del_\rho{} + \frac{1}{\rho^2} \del_\vartheta^2{} + \frac{1}{1 - \rho^2} \del_\varphi^2{} } \ .
\end{equation}
Given the \(U(1) \times U(1)\) symmetry generated by shifts in \(\varphi \) and \(\vartheta\), the eigenfunctions of the Laplacian have the form
\begin{equation}
  \Psi_{Q, \bar Q}(\rho, \vartheta, \varphi) = \Psi_{Q, \bar Q}(\rho) e^{i \bar Q \vartheta} e^{i  Q \varphi} .
\end{equation}
Let us consider the limit where both \(Q\) and \(\bar Q\) are large and of the same order: \(Q = q \Omega\), \(\bar Q = \bar q \Omega\), \(\Omega \gg 1\).
We can use a \ac{wkb}-type argument to show that the eigenvalues of the Laplacian have the form
\begin{equation}
  E = E_0(q, \bar q) \Omega^2 + E_1(q, \bar q) \Omega + E_2 (q, \bar q) + \dots
\end{equation}
The eikonal approximation is the Ansatz
\begin{equation}
  \Psi_{Q, \bar Q} (\rho) = e^{\bar Q S_\vartheta(\rho) +  Q S_\varphi(\rho)} = e^{\Omega \pqty*{\bar q S_\vartheta(\rho) +  q S_\varphi(\rho)}},
\end{equation}
and at leading order in \(\Omega\), the eigenvalue equation for the Laplacian reduces to
\begin{equation}
  \pqty*{ 1 - \rho^2} \pqty*{ \bar q S_\vartheta'(\rho) +  q S_\varphi'(\rho)}^2 - \frac{\bar q^2}{\rho^2} - \frac{ q^2}{1- \rho^2} + 2 k E_0(q, \bar q) = 0 .
\end{equation}
The term proportional to \(q \bar q\) must be \(\rho\)-independent since it has to cancel with a contribution from \(E_0\).
It follows that
\begin{equation}
  S_\vartheta'(\rho)  S_\varphi'(\rho) + \frac{C_1}{ 1 - \rho^2} = 0
\end{equation}
and that
\begin{equation}
  2 k E_0 = C_2 q^2 + C_3 \bar q^2 + 2 C_1 q \bar q.
\end{equation}
In this way we obtain two separate equations for \(S_\varphi(\rho)\): 
\begin{align}
  (S_\varphi')^2 &= \frac{C_1^2 \rho^2}{\pqty*{1 - \rho^2} \pqty*{1 - C_3 \rho^2}} \ , &
  (S_\varphi')^2 &= \frac{1 - C_2 \pqty*{1 - \rho^2}}{\pqty*{1 - \rho^2}^2} \ , 
\end{align}
which are compatible for \(C_1 = C_2 = C_3 = 1\).
The solution of the eikonal equation is then
\begin{equation} 
  \begin{aligned}
    \begin{aligned}
      S_\varphi (\rho) &= \frac{1}{2} \log(1 - \rho^2) \ , \\
      S_\vartheta (\rho) &=  \log(\rho) \ ,    
    \end{aligned} &&&&
    E_0 = \frac{ (q + \bar q)^2}{2k}.
  \end{aligned}
\end{equation}

In fact, in this case it turns out that the eikonal approximation is exact at all orders in \(Q, \bar Q\), and we have
\begin{equation}\label{eq:sectrumS3Lap}
  \frac{1}{2} \Laplacian \Psi_{Q, \bar Q} (\rho,\vartheta, \varphi) + \frac{ \pqty*{Q + \bar Q} \pqty*{ Q + \bar Q + 2 }}{2k} \Psi_{Q, \bar Q} (\rho,\vartheta, \varphi) = 0,
\end{equation}
where
\begin{equation}
  \Psi_{Q, \bar Q} (\rho, \vartheta, \varphi) = \rho^{\bar Q} \pqty*{1 - \rho^2}^{Q/2} e^{i \bar Q \vartheta} e^{i  Q \varphi} = z_1^{\bar Q} z_2^{Q}.
\end{equation}
We will recover this eigenvalue when discussing the exact partition function of the \ac{wzw} model.

\paragraph{Marginal deformation of the three-sphere.}
The $SU(2)_k$ \ac{wzw} admits a continuous line of marginal deformations driven by the current-current operator $J^3 \bar J^3$.   
The  coordinate system that we have introduced above is particularly well adapted to describe this marginal deformation which, in the infinite-\(k\) limit, corresponds to the following background~\cite{Giveon:1993ph,Forste:2003km}:
\begin{align}
  \dd{s}^2 &= k \bqty*{\frac{\dd{\rho}^2}{1- \rho^2} + \frac{\rho^2}{1 + (\lambda^2 -1 ) \rho^2}  \dd{\vartheta}^2 + \frac{\lambda^2\pqty*{1 - \rho^2}}{1 + (\lambda^2 -1 ) \rho^2} \dd{\varphi}^2}, \\
  B&= \frac{k \lambda^2 \rho^2}{1 + (\lambda^2 - 1) \rho^2} \dd{\vartheta} \wedge \dd{\varphi}, \\
  e^{-2 \Phi(\rho)}  &= \frac{\rho}{\sqrt{\det G}} ,
\end{align}
where $\lambda$ is the parameter along the marginal line (the undeformed model has $\lambda=1$). The generalized Laplacian is
\begin{equation}
	\Laplacian_\Phi{} = \Laplacian{} + \frac{\lambda^2 -1}{k} \del^2_\vartheta{} + \frac{1-\lambda^2}{k \lambda^2}\del^2_\varphi{} \ ,
\end{equation}
where $\Laplacian$ is the operator given in Eq.~\eqref{eq:Laplacian}.
Geometrically, the deformation is driven by the \(U(1)\) operators $\del_{\varphi}{}$ and \(\del_{\vartheta}{}\) that commute with the Laplacian.
It follows that the generalized Laplacian admits the same eigenfunctions \( \Psi_{Q, \bar Q} (\rho, \vartheta, \varphi) \) as in the undeformed case, but now with different eigenvalues.
Using again the solution to the eikonal approximation we find:
\begin{gather}
  \frac{1}{2} \Laplacian_{\Phi} \Psi_{Q, \bar Q} (\rho, \vartheta, \varphi) + E(Q, \bar Q) \Psi_{Q, \bar Q} (\rho, \vartheta, \varphi) = 0,\\
  E(Q, \bar Q) = \frac{\pqty*{Q + \bar Q} \pqty*{Q + \bar Q + 2}}{2k} + \frac{1 - \lambda^2 }{2k} \pqty*{ \frac{Q^2}{\lambda^2}  - \bar Q^2 },\label{eq:sectrumDefS3Lap}
\end{gather}
which we will again recover from the partition function in the \(k \to \infty\) limit, see Eq.~\eqref{eq:spectrum_deformed_sphere}.

\section{The SU(2) WZW model at fixed charge}%
\label{sec:wzw}

Up to now, we have studied \acp{cft} that we have assumed to have an \ac{nlsm} description in a certain limit.
However, in two dimensions, some \acp{cft} are exactly solvable.
In these cases, we can directly access the fixed-charge sectors via the partition function. 
This enables us to compare our \ac{nlsm} results and the generic predictions of the large-charge expansion in~\cite{Komargodski:2021zzy} and identify the respective regimes of validity.

Concretely, we will start from the canonical (fixed-charge) partition function on the torus, written as the trace over the states of given charge \(Q\):
\begin{align}
	Z(Q) &= \Tr_{Q} \bqty*{q^{L_0-c/24}\bar q^{\bar L_0-c/24}}, \, & \text{where $q=e^{2\pi i \tau}$,}
\end{align}
and take the cylinder limit,
\begin{align}
	\tau &= i\frac{\beta}{2\pi R}, & \beta&\to\infty,
\end{align}
so that
\begin{equation}
	Z(Q) = \Tr_{Q}\bqty*{e^{-\beta/R(L_0+\bar L_0-c/12)}}.
\end{equation}
From here, we can extract the conformal dimension $\Delta(Q)$ of the lowest operator in the corresponding ensemble as the free energy in the infinite cylinder limit:
\begin{equation}
	-\lim_{\beta\to \infty} \frac{R}{\beta} \log(Z(Q)) = \Delta(Q)  - \frac{c}{12},
\end{equation}
where $\Delta=h+\bar h$ and $h$, $\bar h$ are the conformal weights.

\subsection{The WZW model}
\label{sec:wzw-model}

The simplest non-trivial example of a solvable \ac{cft} in two dimensions is the $SU(2)_k$ \acl{wzw} model, which for integer $k$ is a rational \ac{cft} based on the affine algebra $\widehat{su}(2)_k$.
 In the limit $k\to \infty$ it admits a semi-classical description in terms of an \ac{nlsm} on the three-sphere, which is the group manifold of $SU(2)$.
 Its action is 
\begin{align}
	S&= \frac{k}{16\pi}\int \dd{z}\dd{\bar z}  \Tr[\del^\mu g^{-1}\del_\mu g] + k \Gamma,\\
	\Gamma &=-\frac{i}{24\pi}\int \dd^{3}{y} \epsilon_{\alpha\beta\gamma} \Tr[g^{-1}\del^\alpha g g^{-1} \del^\beta g g^{-1} \del^\gamma g],
\end{align}
where $g$ is an element of $SU(2)$, and the second integral goes over a 3-manifold that has the worldsheet as its boundary. 
The model has a global $SU(2)\times SU(2)$ symmetry since the group can act on the left and on the right. We can thus fix two charges corresponding to a left and a right \(U(1)\). 

For this model, the full partition function is known. To identify the sectors of fixed charge, we start from the grand-canonical partition function which includes the dual chemical potentials:
\begin{equation}
	Z(z, \bar z;q, \bar q) = \Tr[e^{-\beta/R(L_0+\bar L_0-c/12)}y^{J^3_0}\bar y^{\bar J^3_0}],
\end{equation}
where $y=e^{2\pi i z}$ and the $J_0^3$, $\bar J_0^3$ are the Cartan generators of the left and right SU(2). 
This partition function can be expressed in terms of the characters $\chi_l$ of the affine algebra $\widehat{su}(2)$, where $l$ labels the representation:
\begin{equation}
	Z= \sum_{l,l'} \chi_l(z;\tau) M_{ll'} \chi_{l'}(\bar z;\tau).
\end{equation}
We can always choose $M_{ll'}=\delta_{ll'}$.  
The SU(2) \ac{wzw} model has a continuous line of \emph{marginal deformations} which are generated by adding the operator 
\begin{equation}
	\int \dd{z}\dd{\bar z} J^3_0\bar J^3_0
\end{equation}
to the action~\cite{Chaudhuri:1988qb}. In the deformed case, the $SU(2)\times SU(2)$ symmetry of the semi-classical model is broken to $U(1)\times U(1)$.

\subsection{The parafermion decomposition}%
\label{sec:parafermions}

The \ac{wzw} model is made of two building blocks: an $su(2)/u(1)$ piece associated to parafermions and a $u(1)$ associated to a free boson at the self-dual radius.%
\footnote{In stringy terms, this is to say that the radius of the \(U(1)\) is \(\sqrt{k} = \sqrt{\alpha'} \).}
The two pieces are not independent, they are related by an orbifold~\cite{Fateev:1985mm}:
\begin{equation}
  \pqty*{\frac{ su(2)_k}{u(1)}  \otimes u(1)_{\sqrt{k}}}/\setZ_k.
\end{equation}
We intend to fix the charge associated to the bosonic $u(1)_{\sqrt{k}}$, so it is convenient for us to write the characters in terms of this decomposition:
\begin{equation}
	\chi_l(z;\tau) = \Tr_l[q^{L_0-\frac{c}{24}}e^{2\pi i z J_0^3}] = \sum_{m=-k+1}^k c_m^l(q)\theta_{ml}(q,z).
\end{equation}
We use the conventions in which $0\leq l \leq k-1$ is an integer, $-k+1 \leq m \leq k$, and $l-m = 0 \mod 2$.
The theta function is given by
\begin{equation}
	\theta_{ml}(q,z) = \sum_{n \in \mathbb{Z}} q^{l(n + \frac{m}{2l})^2} z^{(ln + \frac{m}{2})}
\end{equation}
and $c_m^l$ are the string functions~\cite{Kac:1984mq} (see also~\cite{DiFrancesco:1997nk}).
The added advantage of this decomposition is that the marginal deformation only acts on the $u(1)$ boson by changing its radius away from the self-dual point (\emph{i.e.} it does not anymore coincide with \(\sqrt{\alpha' }\)), and does not modify the parafermion string functions~\cite{Hassan:1992gi,Yang:1988bi,Giveon:1993ph,Forste:2003km}:
\begin{equation}
	\pqty*{ \frac{ su(2)_k}{u(1)}  \otimes u(1)_{\sqrt{k}\lambda}}/\setZ_k.
\end{equation}
Here $\lambda$ is the deformation parameter,
and $\lambda=1$ corresponds to the undeformed model.
For irrational values of $\lambda^2$, the resulting theory is not a rational \ac{cft}.

Having an explicit expression for the characters, we can derive the full torus partition function for any value of the marginal deformation parameter~\cite{Yang:1988bi}:
\begin{multline}
   Z(z, \bar z;q, \bar q) = \sum_{l=0}^{k-1}\sum_{m=-k+1}^k \sum_{r=0}^{k-1} c_m^l(q) c_{m-2r}^l(\bar q) \sum_{M,N\in \setZ} q^{\frac{1}{4k}(\frac{kM+m-r}{\lambda} + \lambda(kN+r))^2}\\
	\bar q^{\frac{1}{4k}(\frac{kM+m-r}{\lambda} - \lambda(kN+r))^2} y^{kM+m-r}\bar y^{kN+r} \ ,
\end{multline}
which has a manifest \(\lambda \to 1/\lambda \) symmetry, the axial-vector duality~\cite{Gepner:1986hr,Kiritsis:1993ju,Giveon:1993ph}.
The exponents of \(y\) and \(\bar y\) are respectively the charges  $Q$ and $\bar Q$.
It is immediate to specialize to the partition function at fixed charges by imposing
\begin{align}
	Q&=kM+m-r, & \bar Q&=kN+r,
\end{align}
or equivalently,
\begin{align}
  m &= ( Q + \bar Q) + k (M + N), & m - 2r = (Q - \bar Q) + k (M - N).
\end{align}
We now obtain the canonical partition function
\begin{equation}
   Z(Q, \bar Q;q, \bar q) = \sum_{l=0}^{k-1} c_{(Q+\bar Q)_k}^l \bar c_{(Q-\bar Q)_k}^l  q^{\frac{1}{4k}(Q/\lambda + \lambda \bar Q)^2}	\bar q^{\frac{1}{4k}(Q/\lambda - \lambda \bar Q)^2}.
\end{equation}
We are interested in the lowest state with fixed charge.
To find it, first observe that the string functions \(c_m^l\) have the symmetries
\begin{equation}
  c_m^l = c_{-m}^l = c_{k-m}^{k-l} = c_{m+2k}^l  
\end{equation}
and that for \(\abs*{m} \le l\), in the infinite cylinder limit \(q \to 0\)
\begin{equation}
	c_m^l \sim q^{\frac{l(l+2)}{4(k+2)}+\frac{k}{8(k+2)}-\frac{m^2}{4k}}+\dots
\end{equation}
The state of minimal energy is then obtained for the values of \(M\) and \(N\) such that
\begin{equation}
  \begin{cases}
    m = (Q + \bar Q) \mod k \equiv (Q + \bar Q)_k \\
    m - 2 r = (Q - \bar Q) \mod k \equiv (Q - \bar Q)_k \\
  \end{cases}
\end{equation}
and for \(l\) being the smallest value such that \(\abs*{m} \le l\) and \(\abs*{m - 2r} \le l\).
If we assume, without loss of generality, that \(Q > \bar Q > 0\), we have \(l = (Q + \bar Q)_k\), and the free energy gives directly the dimension of the lowest operator, 
\begin{equation}%
 \label{eq:dimension-deformed-WZW}
  \Delta = \frac{(Q+\bar Q)_k \pqty*{(Q + \bar Q)_k + 2}}{2 (k + 2)} - \frac{(Q + \bar Q)_k^2}{4 k } - \frac{(Q - \bar Q)_k^2}{4 k }  + \frac{1}{4k} \pqty*{\frac{Q}{\lambda} + \lambda \bar Q }^2 + \frac{1}{4k} \pqty*{\frac{Q}{\lambda} - \lambda \bar Q }^2.
\end{equation}
This expression is manifestly invariant under the axial-vector duality. %

Depending on the value of the charges there are two qualitatively different behaviors:
\begin{itemize}
\item If \(Q + \bar Q < k\), then \((Q + \bar Q)_k = Q + \bar Q\) and the dimension is
  \begin{equation}\label{eq:spectrum_deformed_sphere}
	\Delta = \frac{(Q+\bar Q)(Q+\bar Q+2)}{2(k+2)} + \frac{1-\lambda^2}{2k}\pqty*{\frac{Q^2}{\lambda^2}-\bar Q^2} \ ,
 \end{equation}
 which, in the special case \(\lambda =1\), reduces to
 \begin{equation}
	\Delta = \frac{(Q+\bar Q)(Q+\bar Q+2)}{2(k+2)}. 
\end{equation}
This is the dimension of a primary along the line of marginal deformations of the  $SU(2)$ \ac{wzw} model parametrized by \(\lambda\).
Geometrically, at the undeformed point \(\lambda=1\) this is the eigenvalue of the Laplacian on a three-sphere of radius $\sqrt{k+2}$
 with angular momentum $Q+\bar Q$. For generic values of \(\lambda\), the $SU(2)$ symmetry is clearly broken, but the expression still only depends on $Q$ and $\bar Q$, and it still can be interpreted as the eigenvalues of a Laplacian on a deformed sphere.
These results reproduce respectively the expressions in Eq.~\eqref{eq:sectrumS3Lap} and  Eq.~\eqref{eq:sectrumDefS3Lap} in the appropriate semiclassical limit \(k \to \infty\).
\item If \(Q + \bar Q > k\), the dimension is
\begin{equation}\label{eq:DimensionSU2BiggerQ}
	\Delta = \frac{1}{2k} \pqty*{\frac{Q^2}{\lambda^2} + \lambda^2 \bar Q^2 }+ a_k(Q, \bar Q),
\end{equation}
where \(a_k(Q, \bar Q)\) is defined in~Eq.\eqref{eq:dimension-deformed-WZW} and is generically of order \(Q^0\) for fixed \(k\).
From the general theory, we expect in this regime a \(U(1) \times U(1)\) sector to decouple.
In fact, the conformal dimension is given by the sum of the contribution of the two fixed charges, that enter precisely with a term proportional to their square (as in the compact free boson discussed in Appendix~\ref{sec:freeBoson}), and another term that is not controlled by the large charge.
The axial-vector duality is then understood as the T-duality that relates the momenta \(Q\) to the windings \(\bar Q\).
In this regime we do not expect the theory to be described by a simple \ac{eft}. The dominating scale is fixed by the large charge \(Q\) which only controls a subsector of the full dynamics. Even for $k$ large, we are not in the standard semi-classical regime of the \ac{wzw} model.
\end{itemize}

Since this model is exactly solvable, we can identify precisely the regimes of validity of the large-charge expansions. If the charge is the dominating controlling parameter ($Q+\bar Q \gg 1$, $Q + \bar Q \gg k$), we see that the two \(U(1)\)s decouple from the rest of the model and the large charge does not control the entire dynamics. The spectrum is the one of a free boson plus order-one corrections.
In the regime where the theory is not controlled by $Q$, but there is an \ac{nlsm} description ($k \gg Q +\bar Q\gg 1$), we find that the scaling dimensions have an expansion in $Q$ and $\bar Q$ starting at $(Q+\bar Q)^2$. In the special \ac{wzw} case we studied, it only contains two terms.

\subsection{Special cases}%
\label{sec:special}

For concreteness, we study some special cases.

\paragraph{k=1.}
At level \(k = 1\), the \(su(2)_1\) model is just a free boson at the self-dual radius and the \(J \bar J\) deformation changes the radius.
Consider~\eqref{eq:dimension-deformed-WZW} with $k=1$.
The function $a_1(Q, \bar Q)$ in Eq.~\eqref{eq:DimensionSU2BiggerQ} vanishes identically for integer \(Q\) and \(\bar Q\), 
and we are left with
\begin{equation}
	\Delta_{FB} =\frac{1}{2} \left[ \frac{Q^2}{\lambda^2} + \lambda^2 \bar{Q}^2 \right].
\end{equation}
This is the lowest operator dimension of the free boson discussed in Appendix~\ref{sec:freeBoson} with both of the $U(1)$ charges fixed, with \(Q\) and \(\bar Q\) identified respectively with momentum and winding number.

\paragraph{k=2.}
For $k=2$, the parafermion decomposition of an \(su(2)_2\) is the orbifold of a free boson and a \(k=2\) parafermion, which is a standard fermion.
The primary fields are given by
\begin{align}
  \Phi_m^l(z) &= \phi_m^l(z) e^{i m \varphi(z)/2}, & \tilde \Phi_{\tilde m}^{\tilde l}(\bar z) &= \tilde \phi_{\tilde m}^{\tilde l}(\bar z) e^{i \tilde m \tilde \varphi(\bar z)/2},  
\end{align}
where the orbifold fixes the values of \(m\) and \(\tilde m\) in the parafermion field to be the same as the \(U(1)\) charge of the boson.
This corresponds to the relations that we had found above:
\begin{align}
  m &= (Q + \bar Q)_2, &  \tilde m &= (Q - \bar Q)_2 \ ,
\end{align}
which in turn fixes the parity of \(l\) and \(\tilde l\), since in general
\begin{align}
  l - m &= 0 \mod 2,  & \tilde l - \tilde m &= 0 \mod 2 .
\end{align}
The parity of \(l\) is related to the boundary conditions on the cylinder:
if \(l\) is odd, we have a \ac{r} \ac{bc}, while
if \(l\) is even, we have a \ac{ns} \ac{bc}.
The dimension of the lowest (Virasoro) primary of charges \((Q, \bar Q)\) will receive two contributions: one from the boson \(U(1)_{\sqrt{2}\lambda}\), the other from the zero-point energy of the appropriate fermionic sector.
\begin{itemize}
\item If \( Q + \bar Q = 0 \mod 2\), we have \acl{ns} \ac{bc} and the free fermion partition function is given by~\cite{DiFrancesco:1997nk}:
  \begin{equation}
    Z_{NS} =  \frac{1}{2} \abs*{\frac{\theta_3}{\eta} } +  \frac{1}{2} \abs*{\frac{\theta_4}{\eta} } = \frac{1}{2 \abs*{\eta}}\pqty*{ \abs*{\sum_{n \in \setZ} q^{n^2/2}} + \abs*{\sum_{n \in \setZ} (-)^{n} q^{n^2/2}} },
  \end{equation}
  where \(q = e^{-\beta/R}\). In the infinite cylinder limit \(\beta \to \infty\), we have
  \begin{equation}
    Z_{NS} \sim  e^{\beta/(24 R)} \pqty*{ 1 + \order{e^{-\beta/(2R)}}}\, ,
  \end{equation}
  where we have used that for small \(q\), \(\eta(q) \sim q^{1/24}\) and that the leading contribution comes from the \(n = 0\) mode.
  It follows that the contribution to the conformal dimension is
  \begin{equation}\label{eq:DeltaNS}
    \Delta_{NS} = - \lim_{\beta \to \infty} \frac{R}{\beta} \log(Z_{NS}) + \frac{c}{12} = -  \frac{1}{24} + \frac{1}{24} = 0 \ .
  \end{equation}
\item If \(Q + \bar Q = 1 \mod 2\), instead, the free fermion partition function with \acl{r} \ac{bc} is
  \begin{equation}
    Z_{R} = \frac{1}{2} \abs*{\frac{\theta_2}{\eta} } = \frac{1}{2 \abs*{\eta}} \abs*{\sum_{r \in \setZ + 1/2} q^{r^{2}/2}}\ .
  \end{equation}
  In this case, in the infinite-cylinder limit the leading contribution comes from \(r = \pm 1/2\) and the partition function becomes
  \begin{equation}
    Z_{R} \sim e^{\beta/(24 R)} e^{-\beta/(8R)}  \pqty*{ 1 + \order{e^{-\beta/R}}} \ .
  \end{equation}
  Then the contribution to the conformal dimension is
  \begin{equation}
    \Delta_{R} = - \frac{1}{24} + \frac{1}{8} + \frac{1}{24} = \frac{1}{8} \, .
  \end{equation}
\end{itemize}

This is to be compared with our result from the general partition function: for \(k = 2\), the function $a_2(Q,\bar Q)$ takes the values  $0$ for $Q$ and $\bar Q$ both even or both odd, and $1/8$ otherwise:
\begin{equation}\label{eq:DeltaR}
	\Delta = \frac{1}{4} \left[ \frac{Q^2}{\lambda^2} + \lambda^2 \bar{Q}^2 \right] + \begin{dcases}
		0 & \text{if \(Q+\bar Q = 0 \mod 2\)}\\
		\frac{1}{8} & \text{if \(Q+\bar Q = 1 \mod 2\)}.
	\end{dcases}
\end{equation}
This is precisely what we found in Eq.~\eqref{eq:DeltaNS} and~\eqref{eq:DeltaR}.

As expected from general arguments, we see that in this case the scaling dimension is made up of the free boson piece which scales as $\order{Q^2, \bar Q^2}$ and a second contribution from the fermion that is independent of the fixed charge.

\section{Conclusion and outlook}%
\label{sec:outlook}

Unlike in higher-dimensional cases, in two-dimensional \acp{cft} we cannot construct an \ac{eft} in terms of an expansion in the charge that controls the dynamics in sectors of large fixed charge~\cite{Komargodski:2021zzy}. Naively, one expects problems from the fact that in two dimensions we do not have Goldstone modes from the spontaneous breaking of the global symmetry, which in higher dimensions serve as the light degrees of freedom in terms of which the \ac{eft} is expressed.
The problem however lies in the fact that if we require unitarity and a discrete spectrum, the \(U(1)\) sector which is controlled by the charge decouples from the full dynamics. If the full dynamics was strongly coupled, the rest of the physics remains perturbatively inaccessible.

The large-charge expansion can however be put to good use when studying a model which per se has an \ac{nlsm} description. Then we can work in a double-scaling limit of large charge in conjunction with the controlling scale of the model. In such a case, we can extract general properties of the spectrum using time-honored approximations such as the \ac{wkb} method. Concretely, we have shown how to compute conformal dimensions via differential equations describing the target space geometry of the \ac{nlsm}.

The results from this method can be verified in the case of fully solvable models such as the $SU(2)_k$ \ac{wzw} model by writing down the full partition function and extracting the fixed-charge sector. 

\bigskip

\noindent{}%
While we have confined ourselves here to the simplest case of the \(SU(2)\) \ac{wzw} model and its marginal $J \bar J$ deformation, it would be interesting to study more complicated solvable models such as the \(SU(N)\) \ac{wzw} model or more general marginal deformations~\cite{Israel:2004vv,Israel:2004cd} with this technique. 
The real merit of working at large charge is that it allows tackling cases which are not solvable but  admit a semi-classical description, as is assumed to be the case for most string theory solutions. We leave these points for future investigation.

\bigskip

\noindent{}%
While the standard large-charge \ac{eft} construction is not directly applicable for two-di\-men\-sion\-al \acp{cft}, we conclude that the approach can be successfully applied in settings admitting a semi-classical description.

\section*{Acknowledgments}

\noindent{}%
{\dosserif We would like to thank Simeon Hellerman and Igor Pesando for illuminating discussions.
The work of S.R. is partly supported by the Swiss National Science Foundation under grant number 200021 192137.
D.O. is partly supported by the \textsc{nccr 51nf40--141869} ``The Mathematics of Physics'' (Swiss\textsc{map}).
}
\appendix
\section{The free boson}
\label{sec:freeBoson}

We consider the \ac{cft} of a free periodic scalar field, $X \simeq X + 2\pi r$,
\begin{equation}
	\mathcal{L} = \frac{1}{2\pi \alpha'} \del X \bar\del X.
\end{equation}
The momentum is quantized,
\begin{align}
	p&=\frac{n}{r}, & n&\in \mathbb{Z},
\end{align}
The second quantum number in the compact case is the winding number $w$,
\begin{align}
	X(\sigma+2\pi) &= X(\sigma) + 2\pi r w, & w&\in \mathbb{Z}.
\end{align}
We can write down the exact partition function for this theory, 
\begin{equation}
	Z= \Tr(q^{L_0-c/24}\bar q^{\bar L_0-c/24}),
\end{equation}
where $q=e^{2\pi i \tau}$ and in our case, $c=1$.
In the cylinder limit, $\tau = i\beta/(2\pi R)$, $\beta\to\infty$, so
\begin{equation}
	Z= \Tr(e^{-\beta/R(L_0+\bar L_0-c/12)}),
\end{equation}
which we can write explicitly in terms of two integers~\cite{DiFrancesco:1997nk}:
\begin{equation}
	Z= \frac{1}{\abs*{\eta(\tau)}^2}\sum_{n,w \in \setZ} \exp*(-\frac{\beta}{2R} \left(\frac{\alpha'n^2}{r^2}+\frac{w^2 r^2}{\alpha'} \right)).
\end{equation}
This system has two U(1) symmetries associated to the quantum numbers $n$ and $w$, and we can fix either of them or both by imposing appropriate boundary conditions.

The conformal dimension $\Delta$ of the lowest operator in a given sector is found via the state-operator correspondence starting from the free energy:
\begin{equation}
	- R \lim_{\beta\to \infty} \frac{1}{\beta} \log Z =\left[\Delta  - \frac{c}{12}\right]
\end{equation}
and $\Delta = L_0 + \bar L_0$.

For $\beta/R \to \infty$, the leading behavior of the Dedekind eta function is
\begin{equation}
	\frac{1}{\abs*{\eta(\tau)}^2} \sim \exp*(\frac{\beta}{12 R}) \ ,
\end{equation}
which measures the contribution of the zero-modes and corresponds to the zero-point energy.
So 
\begin{align}
	Z &\sim e^{\beta/(12 R)} \left[ \sum_{n\in \mathbb{Z}} e^{-\frac{\beta }{2R} \frac{\alpha' }{r^2}n^2 }\sum_{w\in \mathbb{Z}} e^{-\frac{\beta }{2R} \frac{r^2}{\alpha' }w^2}\right].
\end{align}
In this form, the sectors of fixed charge already appear manifestly. 
For fixed $n=Q$,
\begin{align}
	Z &\sim e^{\beta/(12 R)} \left[ e^{-\frac{\beta }{R}  \frac{\alpha' }{2 r^2}Q^2} \sum_{w\in \mathbb{Z}} e^{-\frac{\beta }{R} \frac{r^2}{2 \alpha' }w^2}\right], 
\end{align}
and taking the limit $\beta/R \to \infty$, only $w=0$ survives, so
\begin{equation}
	\Delta_{Q} = \frac{\alpha' }{2r^2}Q^2 + \frac{(c-1)}{12},
\end{equation}
where the second term is zero for the compact boson, \emph{i.e.} the contribution of the central charge cancels the contribution of the Casimir energy.
 
Instead of fixing $n$, we could have fixed $w = \bar Q$, which would have lead to the expression
\begin{equation}
   \Delta_{\bar Q} =  \frac{r^2}{2 \alpha' } \bar Q^2.
\end{equation}
Unsurprisingly, the dimensions of the lowest operators \(\Delta_Q\) and \(\Delta_{\bar Q}\) are related by the T-duality transformation that exchanges \(r^2/\alpha'  \leftrightarrow \alpha'/r^2 \) and swaps \(Q \leftrightarrow \bar Q\).

\setstretch{1}

\printbibliography{}

\end{document}